%Paper: hep-th/9402110
%From: Grosche Christian <I02GRO@DSYIBM.DESY.DE>
%Date: FRI, 18 FEB 94 13:54:37 +0100

%
%----------------------------------------------------------------------
%                       Use PLAIN-TeX
%----------------------------------------------------------------------
\input amstex

%%%%%%%%%%%%%%%%%%%%%%%%%%%%%%%%%%%%%%%%%%%%%%%%%%%%%%%%%%%%%%%%%%%%%
%
% First we have a character check
%
% ! exclamation mark    " double quote
% # hash                ` opening quote (grave)
% & ampersand           ' closing quote (acute)
% $ dollar              % percent
% ( open parenthesis    ) close paren.
% - hyphen              = equals sign
% | vertical bar        ~ tilde
% @ at sign             _ underscore
% { open curly brace    } close curly
% [ open square         ] close square bracket
% + plus sign           ; semi-colon
% * asterisk            : colon
% < open angle bracket  > close angle
% , comma               . full stop
% ? question mark       / forward slash
% \ backslash           ^ circumflex
%
% ABCDEFGHIJKLMNOPQRSTUVWXYZ
% abcdefghijklmnopqrstuvwxyz
% 1234567890
%
%%%%%%%%%%%%%%%%%%%%%%%%%%%%%%%%%%%%%%%%%%%%%%%%%%%%%%%%%%%%%%%%%%%%%

\magnification=1200
\hsize=31pc
\vsize=55 truepc
\hfuzz=2pt
\vfuzz=4pt
\pretolerance=500
\tolerance=500
\parskip=0pt plus 1pt
\parindent=16pt
%

%%%%%%%%%%%%%%%%%%%%%%%%% Font definitions %%%%%%%%%%%%%%%%%%%%%%%%%%
%
% Fonts for title
%
\font\fourteenrm=cmr10 scaled \magstep2
\font\fourteeni=cmmi10 scaled \magstep2
\font\fourteenbf=cmbx10 scaled \magstep2
\font\fourteenit=cmti10 scaled \magstep2
\font\fourteensy=cmsy10 scaled \magstep2

% Font for small caps within title and authors names
%
\font\large=cmbx10 scaled \magstep1

% Font for matrices (please replace with cmbx10
% if you do not have this font available)
%

% Font for matrices (use \bss{x} for bold sans serif within maths)
%

% Font for vectors (bold italic). Please replace with cmbx10
% if you do not have this font available.
% Use \bi{r} for bold italic r within maths
%

% Fonts for small type (if used)
%
\font\eightrm=cmr8
\font\eighti=cmmi8
\font\eightbf=cmbx8
\font\eightit=cmti8

\font\eightsy=cmsy8
\font\sixrm=cmr6
\font\sixi=cmmi6
\font\sixsy=cmsy6

%%%% Definitions of tenpoint, eightpoint and fourteenpoint families
%
\def\tenpoint{\def\rm{\fam0\tenrm}%
  \textfont0=\tenrm \scriptfont0=\sevenrm
                      \scriptscriptfont0=\fiverm
  \textfont1=\teni  \scriptfont1=\seveni
                      \scriptscriptfont1=\fivei
  \textfont2=\tensy \scriptfont2=\sevensy
                      \scriptscriptfont2=\fivesy
  \textfont3=\tenex   \scriptfont3=\tenex
                      \scriptscriptfont3=\tenex
  \textfont\itfam=\tenit  \def\it{\fam\itfam\tenit}%
  \textfont\slfam=\tensl  \def\sl{\fam\slfam\tensl}%
  \textfont\bffam=\tenbf  \scriptfont\bffam=\sevenbf
                            \scriptscriptfont\bffam=\fivebf
                            \def\bf{\fam\bffam\tenbf}%
  \normalbaselineskip=20 truept
  \setbox\strutbox=\hbox{\vrule height14pt depth6pt
width0pt}%
  \let\sc=\eightrm \normalbaselines\rm}
\def\eightpoint{\def\rm{\fam0\eightrm}%
  \textfont0=\eightrm \scriptfont0=\sixrm
                      \scriptscriptfont0=\fiverm
  \textfont1=\eighti  \scriptfont1=\sixi
                      \scriptscriptfont1=\fivei
  \textfont2=\eightsy \scriptfont2=\sixsy
                      \scriptscriptfont2=\fivesy
  \textfont3=\tenex   \scriptfont3=\tenex
                      \scriptscriptfont3=\tenex
  \textfont\itfam=\eightit  \def\it{\fam\itfam\eightit}%
  \textfont\bffam=\eightbf  \def\bf{\fam\bffam\eightbf}%
  \normalbaselineskip=16 truept
  \setbox\strutbox=\hbox{\vrule height11pt depth5pt width0pt}}
\def\fourteenpoint{\def\rm{\fam0\fourteenrm}%
  \textfont0=\fourteenrm \scriptfont0=\tenrm
                      \scriptscriptfont0=\eightrm
  \textfont1=\fourteeni  \scriptfont1=\teni
                      \scriptscriptfont1=\eighti
  \textfont2=\fourteensy \scriptfont2=\tensy
                      \scriptscriptfont2=\eightsy
  \textfont3=\tenex   \scriptfont3=\tenex
                      \scriptscriptfont3=\tenex
  \textfont\itfam=\fourteenit  \def\it{\fam\itfam\fourteenit}%
  \textfont\bffam=\fourteenbf  \scriptfont\bffam=\tenbf
                             \scriptscriptfont\bffam=\eightbf
                             \def\bf{\fam\bffam\fourteenbf}%
  \normalbaselineskip=24 truept
  \setbox\strutbox=\hbox{\vrule height17pt depth7pt width0pt}%
  \let\sc=\tenrm \normalbaselines\rm}
\def\today{\number\day\ \ifcase\month\or
  January\or February\or March\or April\or May\or June\or
  July\or August\or September\or October\or November\or
December\fi
  \space \number\year}
\def\monthyear{\ifcase\month\or
  January\or February\or March\or April\or May\or June\or
  July\or August\or September\or October\or November\or
December\fi
  \space \number\year}

%%%%%%%%%%%%%%%%%%%%%%%% Counter definitions %%%%%%%%%%%%%%%%%%%%%%%%
%
\newcount\secno      %section number
\newcount\subno      %number of subsection
\newcount\subsubno   %number of subsubsection
\newcount\appno      %appendix number
\newcount\tableno    %table number
\newcount\figureno   %figure number
%

%%%%%%%%%%%%%%%%%%%%%%%%%%%%% Baselineskip %%%%%%%%%%%%%%%%%%%%%%%%%%
%
\normalbaselineskip=20 truept
\baselineskip=20 truept

%%%%%%%%%%%%%%%%%%%% Specific formatting commands %%%%%%%%%%%%%%%%%%%
%
% Title of article
%
\def\title#1
   {\vglue1truein
   {\baselineskip=24 truept
    \pretolerance=10000
    \raggedright
    \noindent \fourteenpoint\bf #1\par}
    \vskip1truein minus36pt}
%

% Author names
% (The names of all the authors should be in the form initials then
% surname. There should be no points after initials.)
%
\def\author#1
  {{\pretolerance=10000
    \raggedright
    \noindent {\large #1}\par}}

% Address of the authors
% (If authors are at differing addresses use one \address for each)
%
\def\address#1
   {\bigskip
    \noindent \rm #1\par}

% Short title (not more than fifty characters)
%
\def\shorttitle#1
   {\vfill
    \noindent \rm Short title: {\sl #1}\par
    \medskip}

% Physics Abstracts classification numbers
%
\def\pacs#1
   {\noindent \rm PACS number(s): #1\par
    \medskip}

% Journal article submitted to
%
\def\jnl#1
   {\noindent \rm Submitted to: {\sl #1}\par
    \medskip}

% Today's date
%
\def\date
   {\noindent Date: \today\par
    \medskip}

% Start of abstract
%

% Keyword abstract - only required for J. Phys. G
%
\def\keyword#1
   {\bigskip
    \noindent {\bf Keyword abstract: }\rm#1}

% End of abstract
%

% Contents page (only required for Reports on Progress in Physics)
%
% Heading for contents page
%

% Entry in list of contents (section headings)
%
\def\entry#1#2#3
   {\noindent
    \hangindent=20pt
    \hangafter=1
    \hbox to20pt{#1 \hss}#2\hfill #3\par}

% Subentry in list of contents (subsection heading).
% (Subsubsection headings do not appear in the contents list)
%
\def\subentry#1#2#3
   {\noindent
    \hangindent=40pt
    \hangafter=1
    \hskip20pt\hbox to20pt{#1 \hss}#2\hfill #3\par}
\def\checkforsub{\futurelet\nexttok\decide}
\def\ssf{\relax}
\def\decide{\if\nexttok\ssf\let\endspace=\nospace
                \else\let\endspace=\extraspace\fi\endspace}
\def\nospace{\nobreak\par\nobreak}
%
% Section heading (#1 is title of section, no number required)
%
\def\section#1{%
    \goodbreak
    \vskip24pt plus12pt minus12pt
    \nobreak
    \gdef\extraspace{\nobreak\bigskip\noindent\ignorespaces}%
    \noindent
    \subno=0 \subsubno=0
    \global\advance\secno by 1
    \noindent {\bf \the\secno. #1}\par\checkforsub}

% Subsection heading (#1 is title of subsection, no number required)
%
\def\subsection#1{%
     \goodbreak
     \vskip24pt plus12pt minus6pt
     \nobreak
     \gdef\extraspace{\nobreak\medskip\noindent\ignorespaces}%
     \noindent
     \subsubno=0
     \global\advance\subno by 1
     \noindent {\sl \the\secno.\the\subno. #1\par}\checkforsub}

% Subsubsection heading (#1 is title of subsubsection,
% no number is required)
%
\def\subsubsection#1{%
     \goodbreak
     \vskip15pt plus6pt minus6pt
     \nobreak\noindent
     \global\advance\subsubno by 1
     \noindent {\sl \the\secno.\the\subno.\the\subsubno. #1}\null.
     \ignorespaces}

% Heading for an appendix, #1 is title of appendix,
% no number or letter required
%
\def\appendix#1
   {\vskip0pt plus.1\vsize\penalty-250
    \vskip0pt plus-.1\vsize\vskip24pt plus12pt minus6pt
    \subno=0
    \global\advance\appno by 1
    \noindent {\bf Appendix \the\appno. #1\par}
    \bigskip
    \noindent}

% Heading for subsection within an appendix
%
\def\subappendix#1
   {\vskip-\lastskip
    \vskip36pt plus12pt minus12pt
    \bigbreak
    \global\advance\subno by 1
    \noindent {\sl \the\appno.\the\subno. #1\par}
    \nobreak
    \medskip
    \noindent}

% Heading for acknowledgments
%
\def\ack
   {\vskip-\lastskip
    \vskip36pt plus12pt minus12pt
    \bigbreak
    \noindent{\bf Acknowledgments\par}
    \nobreak
    \bigskip
    \noindent}

%%%%%%%%%%%%%%%%%%%%%%%%%Macros for Tables%%%%%%%%%%%%%%%%%%%%%%%%%%%%

% Heading for start of tables section
%

% Table caption. #1 is caption, no number required
%
\def\tabcaption#1
   {\global\advance\tableno by 1
    \noindent {\bf Table \the\tableno.} \rm#1\par
    \bigskip}

% Definition of boldrule and medrule
%

% The halign (actually ialign) command for tables
% (THIS MUST BE COPIED FOR EACH TABLE---WITHOUT THE PER CENT SIGN!)
%
% \ialign{#\hfil&&\hglue 2pc plus2pc minus1pc#\hfil\cr

% A small negative skip for use in tables
%

% A macro for a footnote to a table
%

% Heading for list of figure captions
%

% Figure caption, #1 is caption, no number required
%
\def\figcaption#1
   {\global\advance\figureno by 1
    \noindent {\bf Figure \the\figureno.} \rm#1\par
    \bigskip}

% Heading for list of references
%

% Heading for list of numbered references
%
\def\numreferences
     {% \vfill\eject
     {\noindent \bf References\par}
     \everypar{\parindent=30pt \hang \noindent}
     \bigskip}

% Reference to a journal article in Harvard (alphabetical) system
%
\def\refjl#1#2#3#4
   {\hangindent=16pt
    \hangafter=1
    \rm #1
   {\frenchspacing\sl #2
    \bf #3}
    #4\par}

% Reference to a book or report in Harvard (alphabetical) system
%
\def\refbk#1#2#3
   {\hangindent=16pt
    \hangafter=1
    \rm #1
   {\frenchspacing\sl #2}
    #3\par}

% Reference to a journal article in numerical system
%
\def\numrefjl#1#2#3#4#5
   {\parindent=40pt
    \hang
    \noindent
    \rm {\hbox to 30truept{\hss #1\quad}}#2
   {\frenchspacing\sl #3\/
    \bf #4}
    #5\par\parindent=16pt}

% Reference to a book or report in numerical system
%
\def\numrefbk#1#2#3#4
   {\parindent=40pt
    \hang
    \noindent
    \rm {\hbox to 30truept{\hss #1\quad}}#2
   {\frenchspacing\sl #3\/}
    #4\par\parindent=16pt}

% Dash for use with repeated authors in reference lists
%

\def\ref#1{\par\noindent \hbox to 21pt{\hss
#1\quad}\frenchspacing\ignorespaces}

% Fraction, alternative to \over
%
\def\frac#1#2{{#1 \over #2}}

% Renaming the dot under macro
%

% \d now used for differential d in mathematics
%

% \e gives roman e for exponential e in mathematics
%
\def\e{\operatorname{e}}
%\def\e{\hbox{\rm e}}

% \i gives roman i for square root of minus one in maths mode
% and \ii used for dotless i in text mode

%\def\i{\hbox{\rm i}}
\def\i{\operatorname{i}}
\chardef\ii="10

% Small (text size) fraction within displayed mathematics
%

% et al
%

% Redefinition of footnote macros to lose rule and remove indentation
%

\catcode`\@=11
\def\vfootnote#1{\insert\footins\bgroup
    \interlinepenalty=\interfootnotelinepenalty
    \splittopskip=\ht\strutbox % top baseline for broken footnotes
    \splitmaxdepth=\dp\strutbox \floatingpenalty=20000
    \leftskip=0pt \rightskip=0pt \spaceskip=0pt \xspaceskip=0pt
    \noindent\eightpoint\rm #1\ \ignorespaces\footstrut\futurelet\next\fo@t}

% Special macros for display equations
%
% \eq(#1) will give the equation number (#1) on the right
% instead of \eqno
%
\def\eq(#1){\hfill\llap{(#1)}}
\catcode`\@=12
%
% Macro for special accented characters
%
% Vectors with hats

% vectors with overbar

% roman characters with right pointing arrow

%
% Abbreviations for IOPP journals
%

        %1968-87
   %1988 and onwards
     %1968--1988
        %1989 and onwards

         %1968-89

           %1975--1988
     %1989 and onwards

         %1989 and onwards

%
% Other commonly quoted journals
%

%
% Miscellaneous definitions
%
% Bold nabla

%
% Bold dot for vector dot products
%

%
% Small space between lines in alignments or displayed maths

%
% Half line space for within tables or alignments

%
% Small negative space to close up lines above rules in tables

%
% greater than approximately signs
% \def\gap{\;\amex{\char'046}\;}           % use these if ams
%  extension fonts available
% \def\sgap{\;\hbox{\samsx \char'046}\;}   % see above
\def\gap{\;\lower3pt\hbox{$\buildrel > \over \sim$}\;}
%
% less than approximately
%
\def\lap{\;\lower3pt\hbox{$\buildrel < \over \sim$}\;}
% space between parts of short equations
\def\tqs{\hbox to 25pt{\hfil}}

   %order of
\def\cal#1{{\Cal #1}}

{\obeylines\gdef\startdisplay#1
  {\catcode`\^^M=5$$#1\halign\bgroup\indent##\hfil&&\qquad##\hfil\cr}}
\outer\def\enddisplay{\crcr\egroup$$}

\chardef\other=12
\def\ttverbatim{\begingroup \catcode`\\=\other \catcode`\{=\other
  \catcode`\}=\other \catcode`\$=\other \catcode`\&=\other
  \catcode`\#=\other \catcode`\%=\other \catcode`\~=\other
  \catcode`\_=\other \catcode`\^=\other
  \obeyspaces \obeylines \tt}
{\obeyspaces\gdef {\ }}  % \obeyspaces now gives \ , not \space

\outer\def\begintt{$$\let\par=\endgraf \ttverbatim \parskip=0pt
  \catcode`\|=0 \rightskip=-5pc \ttfinish}
{\catcode`\|=0 |catcode`|\=\other % | is temporary escape character
  |obeylines % end of line is active
  |gdef|ttfinish#1^^M#2\endtt{#1|vbox{#2}|endgroup$$}}

\catcode`\|=\active
{\obeylines\gdef|{\ttverbatim\spaceskip=.5em plus.25em minus.15em
 \let^^M=\ \let|=\endgroup}}%

\TagsOnRight
\hsize=16.0truecm
\vsize=24.0truecm
\hfuzz=3pt

\tracingstats=1    % Speicherplatzstatistik

\font\twelverm=cmr10 scaled 1200

\normalbaselineskip=12pt
\baselineskip=18pt
%----------------------------------------------------------------------
%                        DEFINITIONS
%----------------------------------------------------------------------
\def\vec#1{{\textfont1=\tenbf\scriptfont1=\sevenbf
\textfont0=\tenbf\scriptfont0=\sevenbf
\mathchoice{\hbox{$\displaystyle#1$}}{\hbox{$\textstyle#1$}}
{\hbox{$\scriptstyle#1$}}{\hbox{$\scriptscriptstyle#1$}}}}
%--------------------------------------------------------------------
\def\bbbr{\operatorname{{I\!R}}}                     %reelle Zahlen
\def\ih{{\i\over\hbar}}
\def\myalign{\allowdisplaybreaks\align}
\def\CD{\cal D}
\def\vKV{\vec K^{(V)}}
\def\vKV{{\vec K}^{(V)}}
\def\vGV{{\vec G}^{(V)}}
\def\vGz{{\vec G}^{(0)}}
\def\GV{G^{(V)}}
\def\hGV{\widehat{G}^{(V)}}
\def\sign{\operatorname{sign}}
\def\erfc{\operatorname{erfc}}

\newcount\glno
\def\plus{\advance\glno by 1}
\def\minus{\advance\glno by -1}
\def\num{\the\glno}

\newcount\Refno
\def\add{\advance\Refno by 1}
\Refno=1

\edef\BADUb{\the\Refno}\add
\edef\CMS{\the\Refno}\add
\edef\CFG{\the\Refno}\add
\edef\GROx{\the\Refno}\add
\edef\FH{\the\Refno}\add
\edef\GBD{\the\Refno}\add
\edef\GROh{\the\Refno}\add
\edef\AGHH{\the\Refno}\add
\edef\GROw{\the\Refno}\add
\edef\JASCH{\the\Refno}\add
\edef\ICHTA{\the\Refno}\add
\edef\DKb{\the\Refno}\add
\edef\GRSb{\the\Refno}\add

{\nopagenumbers
\pageno=0
\centerline{DESY 94 - 019 \hfill ISSN 0418 - 9833}
\centerline{February 1994\hfill}
%\centerline{\hfill hep-th/9402110}
\vskip1cm
\centerline{\fourteenpoint $\delta'$-Function Perturbations and
                           Neumann}
\bigskip
\centerline{\fourteenpoint Boundary-Conditions by Path Integration}
\bigskip
\bigskip
\centerline{\twelverm CHRISTIAN GROSCHE$^*$}
\bigskip
\centerline{\it II.\ Institut f\"ur Theoretische Physik}
\centerline{\it Universit\"at Hamburg, Luruper Chaussee 149}
\centerline{\it 22761 Hamburg, Germany}
\vfill
\midinsert
\narrower
\noindent
{\bf Abstract.}
$\delta'$-function perturbations and Neumann boundary conditions are
incorporated into the path integral formalism. The starting point is
the consideration of the path integral representation for the one
dimensional Dirac particle together with a relativistic point
interaction. The non-relativistic limit yields either a usual
$\delta$-function or a $\delta'$-function perturbation; making their
strengths infinitely repulsive one obtains Dirichlet, respectively
Neumann boundary conditions in the path integral.
\endinsert

\bigskip\noindent
\centerline{\vrule height0.25pt depth0.25pt width4cm\hfill}
\noindent
{\eightpoint\eightrm
 $^*$ Supported by Deutsche Forschungsgemeinschaft under contract
 number GR 1031/2--1.}
\eject}
\pageno=1

%----------------------------------------------------------------------
%                          END OF FILE0
%----------------------------------------------------------------------

\glno=0                      %I
Attempts to incorporate Dirichlet and Neumann boundary conditions into
the path integral formalism are e.g.~due to Barut and Duru$^{\BADUb}$,
Clark et al.$^{\CMS}$ and Carreau et al.$^{\CFG}$. Barut and Duru used a
canonical transformation to Hamilton-Jacobi coordinates in a phase space
path integral to perform the path integration as explicitly as possible
yielding an integral representation of the Feynman kernel; they also
could discuss step potentials within their formalism. In Refs.~[\CMS,
\CFG] general boundary conditions were addressed, but only for the free
particle case.

In a previous paper I have discussed how to implement Dirichlet
boundary conditions into the path integral$^{\GROx}$. This was achieved
by considering a one dimensional $\delta$-function perturbation in the
path integral. This problem can be solved in a straightforward manner by
means of a perturbation expansion$^{\FH,\GBD}$ which can be explicitly
summed yielding the corresponding (energy-dependent) Green function
$G^{(\delta)}(E)$ in terms of the unperturbed one $G^{(V)}(E)$ where $V$
refers to an arbitrary potential which can be included$^\GROh$. Making
the strength of the $\delta$-function infinitely repulsive yields
Dirichlet boundary conditions at the location of the  $\delta$-function
perturbation$^{\AGHH, \GROw}$. It is desireable to have also an
analogous representation for a $\delta'$-function perturbation. Making
in this case the strength of the coupling infinitely repulsive, produces
Neumann boundary conditions at the location of the $\delta'$-function.
However, things turn out to be awkward if one tries to consider by a
similar reasoning as for the usual $\delta$-, a $\delta'$-function
perturbation in the path integral. An expansion into a perturbation
expansion yields interrelated complicated terms with no obvious
resolution of the summation problem. Alternatively, an approximation of
the $\delta'$-function in terms of two usual $\delta$-functions with
distance $\epsilon$ does not make sense in an obvious way. Having this
in mind and that the existing literature concerning $\delta'$-function
perturbations and Neumann boundary conditions in the path integral does
not look satisfactorially, something new is needed and one has to look
for an appropriate regularization procedure to fill the gap.

In this paper this problem is resolved by means of the path integral
representation of the one dimensional Dirac particle$^{\FH}$. The
incorporation of a point-interaction yields a two parameter family for
the corresponding self-adjoint extension$^{\AGHH}$: One can choose
either the up (or electron component) or the down (or positron)
component where the point-interaction is acting on. Considering a
perturbation expansion for both problems, it is found that they can be
explicitly summed in terms of the corresponding Green functions (which
are a $2\times2$ matrices). In the non-relativistic limit the former
case yields the usual $\delta$-function perturbation, whereas in the
latter we obtain the equivalent of a $\delta'$-function perturbation
in the path integral. We will concentrate on this case.

In the following I will outline how to implement point-interactions
in the one dimensional Dirac particle path integral. We obtain in
the case of the $\delta'$-function perturbation automatically a correct
regularization prescription in terms of the unperturbed Green function
$G^{(V)}(E)$. The general method for the time-ordered perturbation
expansion is quite simple. We assume that we have a potential
$W(x)=V(x)+\widetilde V(x)$ in the path integral and we suppose that
$W$ is so complicated that a direct path integration is not possible.
However, the path integral $K^{(V)}$ corresponding to $V(x)$ is assumed
to be known. We expand the path integral containing $\widetilde V(x)$
in a perturbation expansion about $V(x)$ in the following way. The
initial kernel corresponding to $V$ propagates in $\Delta t$-time
unperturbed, then it interacts with $\widetilde V$, propagates again in
another $\Delta t$-time unperturbed, a.s.o, up to the final state. This
gives the series expansion$^{\FH,\GBD}$ ($x\in\bbbr$)
\plus
$$\myalign
  K&(x'',x';T)
  =K^{(V)}(x'',x';T)+\sum_{n=1}^\infty\bigg(-\ih\bigg)^n
  \left(\prod_{j=1}^n
  \int_{t'}^{t_{j+1}} dt_j\int_{-\infty}^\infty dx_j\right)
  \\   &\times K^{(V)}(x_1,x';t_1-t')
  \widetilde V(x_1)K^{(V)}(x_2,x_1;t_2-t_1)
  \times\dotsc
  \\   &\dots\times
  \widetilde V(x_{n-1})K^{(V)}(x_n,x_{n-1};t_n-t_{n-1})
  \widetilde V(x_n)K^{(V)}(x'',x_n;t''-t_n)\enspace.
  \tag\num\endalign$$
I have ordered time as $t'=t_0<t_1<t_2<\dots<t_{n+1}=t''$ and paid
attention to the fact that $K(t_j-t_{j-1})$ is different from zero only
if $t_j>t_{j-1}$. We consider the path integral representation for the
one dimensional Dirac equation$^{\FH, \JASCH, \ICHTA}$
($p_x=-\i\hbar\partial_x$)
\plus$$\myalign
  \vKV(x'',x';T)
  &=\bigg<x''\bigg\vert\exp\bigg[-\ih T\Big(c\sigma_xp_x+mc^2\sigma_z
           +\vec V(x)\Big)\bigg]\bigg\vert x'\bigg>
        \\   &
  =\int\limits_{x(t')=x'}^{x(t'')=x''}\CD\nu(t)
    \exp\Bigg(-\ih\int_{t'}^{t''}\vec V(x)dt\Bigg)\enspace.
  \tag\num\endalign$$
\edef\numf{\num}%
$\vec V$ may be a matrix-valued potential. The support property of the
measure $\CD\nu$ is defined in such a way that the motion it is
describing selects paths of $N$ steps each of length $c\epsilon$
($\epsilon=T/N$ in the lattice representation) that start at $x'$ in
the direction $\alpha$, and end at $x''$ in the direction $\beta$, where
 $\alpha$ and $\beta$ take the values ``right'' and ``left''. The path
integration then is a summation over all reversings of directions$^\FH$.
$\sigma_x,\sigma_y,\sigma_z$ are the Pauli matrices. We introduce the
Green function $\vGV(E)$ with its matrix representation
\plus$$
   \vGV(x'',x';E)=\pmatrix
    \GV_{11}(x'',x';E)  &\GV_{12}(x'',x';E)   \\
    \GV_{21}(x'',x';E)  &\GV_{22}(x'',x';E)\endpmatrix\enspace.
  \tag\num$$
\edef\numa{\num}%
We first consider a $\delta$-function perturbation in the electron
($=\hbox{``$+$''}$-) component, i.e.\ $\tilde{\vec V}=-\alpha\pmatrix1&0
\\0&0\endpmatrix\delta(x-a)$. We obtain by inserting it into the path
integral and summing the perturbation expansion
\plus$$\myalign
  &{\vec G}^{(\delta_+)}(x'',x';E)
  =\vGV(x'',x';E)+{1\over1/\alpha-G^{(V)}_{11}(a,a;E)}
        \\   &\qquad\times
   \pmatrix
   G^{(V)}_{11}(a,x';E)G^{(V)}_{11}(x'',a;E)
  &G^{(V)}_{11}(a,x';E)G^{(V)}_{12}(x'',a;E)  \\
   G^{(V)}_{21}(a,x';E)G^{(V)}_{11}(x'',a;E)
  &G^{(V)}_{21}(a,x';E)G^{(V)}_{12}(x'',a;E)  \endpmatrix\enspace.
  \tag\num\endalign$$
Similarly for the positron ($=\hbox{``$-$''}$-) component, i.e.\
$\tilde{\vec V}=(4m^2\beta c^2/\hbar^2)\pmatrix0&0\\0&1\endpmatrix
\delta (x-a)$ (the constants have been chosen for convenience)
\plus$$\myalign
  &{\vec G}^{(\delta_-)}(x'',x';E)
  =\vGV(x'',x';E)-{1\over \hbar^2/4m^2c^2\beta+G^{(V)}_{22}(a,a;E)}
        \\   &\qquad\times
   \pmatrix
   G^{(V)}_{12}(a,x';E)G^{(V)}_{21}(x'',a;E)
  &G^{(V)}_{12}(a,x';E)G^{(V)}_{22}(x'',a;E)  \\
   G^{(V)}_{22}(a,x';E)G^{(V)}_{21}(x'',a;E)
  &G^{(V)}_{22}(a,x';E)G^{(V)}_{22}(x'',a;E)  \endpmatrix\enspace.
  \tag\num\endalign$$
\edef\numc{\num}%
We consider the unperturbed free particle; the explicit expression for
$\vGz(E)$ has the form$^{\AGHH}$:
\plus$$\vGz(x'',x';E)
  ={\i\over2c\hbar}\pmatrix
   \zeta          &\sign(x''-x')   \\
   \sign(x''-x')  &1/\zeta         \endpmatrix
   \e^{\i k\vert x''-x'\vert }\enspace,
  \tag\num$$
where $\zeta=(E+mc^2)/ck\hbar$, $ck\hbar=\sqrt{E^2-m^2c^4}$. This
yields for a $\delta$-function perturbation in the electron component:
\plus$$\myalign
  &{\vec G}^{(\delta_+)}(x'',x';E)
  ={\i\over2c\hbar}\pmatrix
   \zeta          &\sign(x''-x')   \\
   \sign(x''-x')  &1/\zeta  \endpmatrix \e^{\i k\vert x''-x'\vert }
        \\   &\qquad
  -{\alpha\e^{\i k(\vert x''-a\vert +\vert a-x'\vert )}\over
    4c\hbar(c\hbar-\i\alpha\zeta/2)}
   \pmatrix
   \zeta^2           &\zeta\sign(x''-a)       \\
   \zeta\sign(a-x')  &\sign(x''-a)\sign(a-x') \endpmatrix\enspace.
  \tag\num\endalign$$
For $[\alpha]>0$ there is one bound state with energy $E=mc^2(1-\lambda^
2)/(1+\lambda^2)$ ($\lambda=\alpha/2c\hbar$). Similarly for a
$\delta$-function perturbation in the positron component
\plus$$\myalign
  &{\vec G}^{(\delta_-)}(x'',x';E)
  ={\i\over2c\hbar}\pmatrix
   \zeta          &\sign(x''-x')   \\
   \sign(x''-x')  &1/\zeta  \endpmatrix \e^{\i k\vert x''-x'\vert }
        \\   &\qquad
  +{2m^2\beta\e^{\i k(\vert x''-a\vert +\vert a-x'\vert )}\over
    \hbar(2\hbar^3+4\i m^2c\beta/\zeta)}
   \pmatrix
   \sign(x''-a)\sign(a-x')  &\sign(a-x')/\zeta  \\
   \sign(x''-a)/\zeta       &1/\zeta^2          \endpmatrix\enspace.
  \tag\num\endalign$$
For $[\beta]>0$ there is one bound state with energy $E=-mc^2(1-\lambda^
2)/(1+\lambda^2)$ ($\lambda=2m^2c\beta/\hbar^3$). Let us assume for
simplicity that the component $\GV_{11}(E)$ in (\numa) is known and
$\vec V$ is a scalar, we then can derive
$$\myalign
  \GV_{12}(x,y;E)
  &={c\over mc^2+V+E}p_x \GV_{11}(x,y;E)\enspace,
  \tag\num\\   \global\plus
  \GV_{22}(x,y;E)
  &={-1\over mc^2+V+E}\Bigg(
    {c^2\over mc^2+V+E}p_xp_y\GV_{11}(x,y;E)+\delta(x-y)\Bigg)\enspace.
  \tag\num\endalign$$
\edef\numb{\num}%
{}From these representations it is easily seen that if $\GV_{11}(E)$
is of $O(1)$ for $c\to\infty$, $\GV_{12}(E)$ and $\GV_{22}(E)$
vanish according to $\propto1/c$ and $\propto1/c^2$ for $c\to\infty$,
respectively.

We consider the limit $c\to\infty$ in ${\vec G}^{(\delta_\pm)}(E)$. On
the one hand we know that the path integral representation (\numf)
gives the usual one dimensional path integral in non-relativistic
quantum mechanics$^{\FH}$. On the other we find that only the $(1,1)$
component in the Green functions remains finite, all others vanish.
Furthermore we have $G_{11}^{(\delta_+)}(E)\to G^{(\delta)}(E)$ and
$G_{11}^{(\delta_-)}(E)\to G^{(\delta')}(E)$, where $G^{(\delta)}(E)
$ is the Green function for a potential problem $V$ with a usual
$\delta$-function perturbation in non-relativistic quantum mechanics,
and $G^{(\delta')}(E)$ is the Green function for a potential problem
$V$ together with a $\delta'$-function perturbation, respectively.
Putting everything together we obtain for the latter an explicit path
integral representation yielding
\minus$$\align
  &G^{(\delta')}(x'',x';E)
         \\   &
  =\ih        \int_0^\infty dT\,\e^{\i ET/\hbar}
  \int\limits_{x(t')=x'}^{x(t'')=x''}\CD x(t)
  \exp\left\{\ih\int_{t'}^{t''}\bigg[{m\over2}\dot x^2
       -V(x)+\beta\delta'(x-a)\bigg]dt\right\}\qquad
  \tag\num\\   \global\plus
              &
  =G^{(V)}(x'',x';E)
   -\dsize\thickfrac{G^{(V)}_{,x'}(x'',a;E)G^{(V)}_{,x''}(a,x';E)}
   {\hGV_{,x'x''}(a, a;E)+1/\beta}\enspace,
  \tag\num\\   \global\plus
  &\hGV_{,xy}(a, a;E)
  =\Big[\partial_x\partial_y G^{(V)}(x,y;E)
      -2m\delta(x-y)/\hbar^2\Big]\bigg\vert_{x=y=a}\enspace.
  \tag\num\endalign$$
\minus\minus
\edef\numd{\num}\plus\plus%
The path integral (\numd) has been {\it derived\/} in a unique way
through the regularization (\numc) in the limit $c\to\infty$. Note that
(\numb) yields automatically the correct regularization of the formal
expression ``$G^{(V)}_{,xy}(a,a;E) $''. For $V\equiv0$, i.e.\ the free
particle, we obtain the explicit representation (together with an
inverse Laplace-Fourier transformation)
\plus$$\myalign
  &\int\limits_{x(t')=x'}^{x(t'')=x''}\CD x(t)
  \exp\Bigg\{\ih\int_{t'}^{t''}
  \bigg[{m\over2}\dot x^2+\beta\delta'(x-a)\bigg]dt\Bigg\}
         \\   &
  =\sqrt{m\over2\pi\i\hbar T}
  \exp\bigg({\i m\over2\hbar T}\vert x''-x'\vert ^2\bigg)
         \\   &\qquad
  +\sqrt{m\over2\pi\i\hbar T}
   \exp\bigg[{\i m\over2\hbar T}(\vert x''-a\vert+\vert x'-a\vert)^2
  \bigg]\sign(x''-a)\sign(x'-a)
         \\   &\qquad
  +{\hbar^2\over2m\beta}\exp\Bigg[-{\hbar^2\over m\beta}
        \Big(\vert x''-a\vert +\vert x'-a\vert \Big)
        +\ih{\hbar^6\over2m^3\beta^2}T\Bigg]
         \\   &\qquad\qquad\times
  \erfc\Bigg\{\sqrt{m\over2\i\hbar T}\,
       \bigg[\Big(\vert x''-a\vert +\vert x'-a\vert \Big)
       -{\i\hbar^3 T\over m^2\beta}\bigg]\Bigg\}
  \sign(x''-a)\sign(x'-a)\enspace.
  \tag\num\endalign$$
The one bound state has energy $E=-\hbar^6/2m^3\beta^2$.
Repeating the procedure for N-fold $\delta'$-function perturbations
gives similarly (compare also Ref.~[\AGHH], $a_i\not= a_j$ ($i\not=j$)
\goodbreak\noindent
\plus$$\myalign
  & \ih \int_0^\infty  dT\,\e^{\i TE/\hbar}
  \int\limits_{x(t')=x'}^{x(t'')=x''}\CD x(t)\exp\left\{\ih
   \int_{t'}^{t''}\left[{m\over2}\dot x^2-V(x)
   +\sum_{j=1}^N\beta_j\delta'(x-a_j)\right]dt\right\}
         \\   &
  =\dsize\thickfrac{\left\vert\matrix
  \GV(x'',x';E)  &\GV_{,x'}(x'',a_1;E) &\hdots
                                 &\GV_{,x'}(x'',a_N;E)       \\
  \GV_{,x''}(a_1,x';E)
                 &\hGV_{,x'x''}(a_1,a_1;E)+1/\beta_1
                 &\hdots         &\GV_{,x'x''}(a_1,a_N;E)    \\
  \vdots         &\vdots         &\ddots            &\vdots  \\
  \GV_{,x''}(a_N,x';E)  &\GV_{,x'x''}(a_N,a_1)   &\hdots
                 &\hGV(a_N,a_N;E)+1/\beta_N
  \endmatrix\right\vert}{\left\vert\matrix
  \hGV_{,x'x''}(a_1,a_1;E)+1/\beta_1
                 &\hdots         &\GV_{,x'x''}(a_1,a_N;E)    \\
  \vdots         &\ddots         &\vdots                     \\
  \GV_{,x'x''}(a_N,a_1;E)        &\hdots
                 &\hGV_{,x'x''}(a_N,a_N;E)+1/\beta_N
  \endmatrix\right\vert}\enspace.
  \tag\num\endalign$$
Of course, any combination of N-fold $\delta$-functions and M-fold
$\delta'$-functions perturbations is possible yielding a closed
expression in terms of the corresponding Green functions.

Making now the coupling of the $\delta'$-function perturbation
infinitely repulsive produces Neumann boundary conditions at $x=a$, i.e.
\plus$$\myalign
  &\ih        \int_0^\infty dT\,\e^{\i ET/\hbar}
  \int\limits_{x(t')=x'}^{x(t'')=x''}\CD_{x=a}^{(N)} x(t)
  \exp\left\{\ih\int_{t'}^{t''}\bigg[{m\over2}\dot x^2
       -V(x)\bigg]dt\right\}
         \\   &
  =G^{(V)}(x'',x';E)
   -G^{(V)}_{,x'}(x'',a;E)G^{(V)}_{,x''}(a,x';E)
   \Big/\hGV_{,x'x''}(a, a;E)\enspace.
  \tag\num\endalign$$
The notation $\CD_{x=a}^{(N)}$ stands for Neumann boundary
conditions at $x=a$, and for the corresponding Green function we write
shorthand $G^{(V,N)}_{x=a}(E)$. Note that $\lim_{\gamma\to-\infty}
G^{(\delta)}(E)=G^{(V,D)}_{x=a}(E)$, where $D$ stands for Dirichlet
boundary conditions$^{\GROw}$. Let us consider a particle moving under
the influence of a potential in the box $a<x<b$ with Neumann boundary
conditions for $x=a$ and $x=b$; we obtain
\plus$$\myalign
   \ih                    \int_0^\infty dT\,&\e^{\i ET/\hbar}
  \int\limits_{x(t')=x'}^{x(t'')=x''}\CD_{(a<x<b)}^{(NN)}x(t)
  \exp\left\{\ih\int_{t'}^{t''}\bigg[{m\over2}\dot x^2
       -V(x)\bigg]dt\right\}
         \\   &
  =\dsize\thickfrac{\left\vert
  \matrix\GV(x'',x';E)      &\GV_{,x'}(x'',b;E)  &\GV_{,x'}(x'',a;E)  \\
         \GV_{,x''}(b,x';E) &\hGV_{,x'x''}(b,b;E)&\GV_{,x'x''}(b,a;E)\\
         \GV_{,x''}(a,x';E) &\GV_{,x'x''}(a,b;E)&\hGV_{,x'x''}(a,a;E)
  \endmatrix\right\vert}{\left\vert\matrix
  \hGV_{,x'x''}(b,b;E) &\GV_{,x'x''}(b,a;E)\\
  \GV_{,x'x''}(a,b;E) &\hGV_{,x'x''}(a,a;E)\endmatrix\right\vert}
  \enspace.
  \tag\num\endalign$$
Of course, any combination of boundary conditions of a particle moving
in the box $a<x<b$ is allowed yielding closed expression in terms of
the corresponding Green functions.

It is now obvious how to describe potential problems with absolute
value dependence, i.e.~$V=V(\vert x\vert)$. Combing the results for
Dirichlet and Neumann boundary conditions we obtain the general formula
\plus$$\myalign
  &\ih                    \int_0^\infty dT\,\e^{\i ET/\hbar}
  \int\limits_{x(t')=x'}^{x(t'')=x''}\CD x(t)
  \exp\left\{\ih\int_{t'}^{t''}\bigg[{m\over2}\dot x^2
       -V(\vert x\vert )\bigg]dt\right\}
         \\   &
  =G^{(V)}(x'',x';E)-{1\over2}
  \Big[G^{(V,D)}_{x=0}(x'',x';E)+G^{(V,N)}_{x=0}(x'',x';E)\Big]\enspace.
  \tag\num\endalign$$
\edef\nume{\num}%
If the potential $V$ already contains only even powers in $x$, like the
harmonic oscillator, the last two term in (\nume) cancel. Simple
examples for the general case are, for instance, the double-oscillator
$V(x)={m\over2}\omega^2(\vert x\vert-a)^2$, the one dimensional
Coulomb-problem $V(x)=k\vert x\vert$, or the symmetric potential well.
For the one dimensional Coulomb problem one obtains e.g.\ the Green
function
\plus$$\myalign
  &\ih                    \int_0^\infty dT\,\e^{\i ET/\hbar}
   \int\limits_{x(t')=x'}^{x(t'')=x''}\CD x(t)
  \exp\left[\ih\int_{t'}^{t''}\bigg({m\over2}\dot x^2
       -k\vert x\vert\bigg)dt\right]
         \\   &
  ={4\over3}{m\over\hbar^2}\bigg[
   \bigg(x'-{E\over k}\bigg)\bigg(x''-{E\over k}\bigg)\bigg]^{1/2}
         \\   &\quad\times\Bigg\{
   K_{1/3}\left({2\over3}{\sqrt{2mk}\over\hbar}
   \bigg(x_>-{E\over k}\bigg)^{3/2}\right)
   I_{1/3}\left({2\over3}{\sqrt{2mk}\over\hbar}
   \bigg(x_<-{E\over k}\bigg)^{3/2}\right)
         \\   &\qquad-{1\over2}
   K_{1/3}\left({2\over3}{\sqrt{2mk}\over\hbar}
   \bigg(x'-{E\over k}\bigg)^{3/2}\right)
   K_{1/3}\left({2\over3}{\sqrt{2mk}\over\hbar}
   \bigg(x''-{E\over k}\bigg)^{3/2}\right)
         \\   &\qquad\quad\times\Bigg[
   I_{1/3}\left({2\over3}{\sqrt{2mk}\over\hbar}
   \bigg(-\dsize{E\over k}\bigg)^{3/2}\right)\Bigg/
   K_{1/3}\left({2\over3}{\sqrt{2mk}\over\hbar}
   \bigg(-\dsize{E\over k}\bigg)^{3/2}\right)
         \\   &\qquad\qquad-2\pi
   I_{2/3}\left({2\over3}{\sqrt{2mk}\over\hbar}
   \bigg(-\dsize{E\over k}\bigg)^{3/2}\right)\Bigg/
   K_{2/3}\left({2\over3}{\sqrt{2mk}\over\hbar}
   \bigg(-\dsize{E\over k}\bigg)^{3/2}\right)\Bigg]\Bigg\}\enspace,
  \tag\num\endalign$$
with the quantization condition
\plus$$
   K_\nu\left({2\over3}{\sqrt{2mk}\over\hbar}
   \bigg(-\dsize{E_n\over k}\bigg)^{3/2}\right)=0\enspace,
  \tag\num$$
with $n={1\over3},{2\over3}$ for the odd, respectively even
wave functions.

In this paper I have presented a perturbation expansion approach to the
problem of $\delta'$-function perturbations and Neumann boundary
conditions in the context of path integrals. This was achieved by
considering the path integral representation of the one dimensional
Dirac particle with a $\delta$-function perturbation in the positron
component. In the non-relativistic limit a $\delta'$-function
perturbation in the path integral emerges. I obtained closed formul\ae\
for both problems in terms of the corresponding energy-dependent Green
function. An analogous  discussion for the electron-component yields a
$\delta$-function perturbation in the path integral and Dirichlet
boundary condition, respectively. The formalism can be repeated in an
obvious way to incorporate multiple $\delta$- and $\delta'$-function
perturbations, and one can consider motion in a box $a<x<b$ with any
combination of Dirichlet and Neumann boundary conditions at the walls of
the box. Analogously to Ref.~[\GROw] one can also generalize our method
to higher dimensions to derive path integral formulations for
$\delta'$-function perturbations, Dirichlet and Neumann boundary
conditions along lines and hyperplanes, etc., respectively.

I could also derive a general expression for potentials with absolute
value dependence by combining the results from Dirichlet and Neumann
boundary conditions, c.f.~(\nume). In general, only the corresponding
Green function can be stated.

The definition of the path integral of the $\delta'$-function
perturbation and its (energy dependent) Green function via the path
integral representation of the one dimensional Dirac particle
looks at first sight circumstancial. However, specific regularization
prescriptions of singular potentials are familiar for path integrals:
For instance, the $1/r$ potential requires in a proper path integral
representation a regularization through the Kustaanheimo-Stiefel
transformation$^{\DKb}$, and the $1/r^2$ potential by means of the
Besselian functional weight$^{\GRSb}$. In the path integral formulation
the usual $\delta$-function perturbation is quite a simple
object$^{\GROh}$ in comparison to the $\delta'$-function perturbation
as shown in this paper. Actually both point interactions describe a
particular kind of boundary conditions of the wave functions in their
domains at the location of the interaction. The even more singular two
and three dimensional point interactions require also a regularization
prescription by means of their Green functions$^{\AGHH}$.

The achieved results for a proper approach to $\delta$-and
$\delta'$-function perturbations, and Dirichlet and Neumann boundary
conditions respectively, in the language of Feynman path integrals
cover properly combined a wide range of problems in path integral
technique. What remains is to develop a path integral formalism to
incorporate general boundary conditions, where Dirichlet and Neumann
boundary conditions are but special cases, respectively multiple
boundary conditions on the real line (e.g.\ combination of step
potentials). These open question will be subject to future
investigations.

\ack
I would like to thank C.\ Oldhoff for fruitful discussions.

\newpage\noindent
\numreferences
%-----------------------------------------------------------------------
\ref{$^{\BADUb}$}
A.O.Barut and I.H.Duru:
Path Integration Via Hamilton-Jacobi Coordinates and Applications to
Potential Barriers;
{\it Phys.Rev.}\ {\bf A 38} (1988) 5906
%-----------------------------------------------------------------------
\ref{$^{\CMS}$}
T.E.Clark, R.Menikoff and D.H.Sharp:
Quantum Mechanics on the Half-Line Using Path Integrals;
{\it Phys.Rev.}\ {\bf D 22} (1980) 3012
%-----------------------------------------------------------------------
\ref{$^{\CFG}$}
M.Carreau, E.Farhi and S.Gutmann:
Functional Integral for a Free Particle in a Box;
{\it Phys.Rev.}\ {\bf D 42} (1990) 1194
%-----------------------------------------------------------------------
\ref{$^{\GROx}$}
C.Grosche:
Path Integration via Summation of Perturbation Expansions and
Application to Totally Reflecting Boundaries and Potential Steps;
{\it Phys.Rev.Lett.}\ {\bf 71} (1993) 1
%-----------------------------------------------------------------------
\ref{$^{\FH}$}
R.P.Feynman and A.Hibbs: Quantum Mechanics and Path Integrals
({\it McGraw Hill}, New York, 1965)
%-----------------------------------------------------------------------
\ref{$^{\GBD}$}
M.J.Goovaerts, A.Babcenco and J.T.Devreese:
A New Expansion Method in the Feynman Path Integral Formalism:
Application to a One-Dimensional Delta-Function Potential;
{\it J.Math.Phys.}\ {\bf 14} (1973) 554
%-----------------------------------------------------------------------
\ref{$^{\GROh}$}
C.Grosche:
Path Integrals for Potential Problems With $\delta$-Function
Perturbation;
{\it J.Phys.A: Math.Gen.}\ {\bf 23} (1990) 5205
%-----------------------------------------------------------------------
\ref{$^{\AGHH}$}
S.Albeverio, F.Gesztesy, R.J.H\o egh-Krohn and H.Holden:
Solvable Models in Quantum Mechanics
({\it Springer}, Berlin-Heidelberg, 1988)
%-----------------------------------------------------------------------
\ref{$^{\GROw}$}
C.Grosche:
$\delta$-Function Perturbations and Boundary Problems by Path
Integration;
{\it Ann.Physik} {\bf 2} (1993) 557
%-----------------------------------------------------------------------
\ref{$^{\JASCH}$}
T.Jacobson and L.S.Schulman:
Quantum Stochastics: The Passage from a Relativistic to a
Non-Relativistic Path Integral;
{\it J.Phys.A: Math.Gen.}\ {\bf 17} (1984) 375
%-----------------------------------------------------------------------
\ref{$^{\ICHTA}$}
T.Ichinose and H.Tamura:
Path Integral Approach to Relativistic Quantum Mechanics;
{\it Prog.Theor.Phys.Supp.}\ {\bf 92} (1987) 144
%-----------------------------------------------------------------------
\ref{$^{\DKb}$}
I.H.Duru and H.Kleinert:
Quantum Mechanics of H-Atoms From Path Integrals;
{\it Fort\-schr.Phys.}\ {\bf 30} (1982) 401
%-----------------------------------------------------------------------
\ref{$^{\GRSb}$}
C.Grosche and F.Steiner:
Path Integrals on Curved Manifolds;
{\it Zeitschr.Phys.}\ {\bf C 36} (1987) 699
%-----------------------------------------------------------------------

\enddocument